\documentclass[preprint,pre,aps,showpacs]{revtex4}
\usepackage{graphicx}
\begin{document}

\title{Cross-sections of neutron rich nuclei  from projectile fragmentation: 
canonical thermodynamic model estimates}
 
\author{G. Chaudhuri \footnote{gargi@physics.mcgill.ca} \footnote{On leave from Variable Energy Cyclotron Center, 1/AF Bidhan Nagar, Kolkata 700064, India} $^1$, S. Das Gupta$^1$, W. G. Lynch$^2$, M. Mocko$^2$,
and M. B. Tsang$^2$}

\affiliation{$^1$Physics Department, McGill University, 
Montr{\'e}al, Canada H3A 2T8}

\affiliation{$^2$National Superconducting Cyclotron Laboratory,Physics $\&$
Astronomy Department, Michigan State University, East Lansing,
 Michigan 48824, USA}

\date{\today}

\begin{abstract}

A remarkably simple dependence of fragmentation cross-section on average
binding energy has been established in experimental data.  This 
dependence was empirically parametrised leading to a very useful formula for
extrapolation.  We find that the canonical thermodynamic model, which
has been used in the past for successful computations of many 
observables resulting from multifragmentation, reproduces the salient
fetures of fragmentation cross-sections of very neutron rich nuclei 
very well.  This helps towards a theoretical understanding of the
observed data.

\end{abstract}


\maketitle

As part of a drive towards understanding the production mechanisms 
of rare isotopes,
fragmentation cross-sections of many neutron rich isotopes have recently
been measured from the $^{48}$Ca and $^{64}$Ni beams at 140 MeV per
nucleon on $^9$Be and $^{181}$Ta targets \cite{Mocko1}.  
 Copper isotope cross-sections 
 have been measured in projectile fragmentation of $^{86}$Kr at
64 MeV per nucleon \cite{Mocko2}.
A remarkable feature is the correlation between
the measured fragment cross-section and the average binding energy
(Fig.1)\cite{Mocko3}.  This observation has prompted attempts of paramtrisation
of cross-sections.  One very successful parametrisation\cite{Mocko3} is
\begin{eqnarray}
\sigma=C\exp (A^{-1}(B-\varepsilon_{pair})/\tau)
\end{eqnarray}
Here $B$ is the binding energy of the nucleus with mass number $A$,
$\varepsilon_{pair}=\kappa\varepsilon A^{-3/4}$ and $\kappa$
is 1 for even-even nuclei, 0 for odd-even nuclei and -1 for odd-odd
nuclei.  The pairing term smooths the straggling seen in the data
when logarithms of cross-sections are plotted against $B/A$ for
even-even and odd nuclei (or odd-even and odd-odd nuclei).  Theoretical
basis for the simple appearence of $B/A$ or the pairing term correction
is not transparent.
Another equation which is highly successful is
\begin{eqnarray}
\sigma=cA^{3/2}\exp [(N\mu_n+Z\mu_p+B-\varepsilon_{pair}+E_s.min(S_n,S_p,
S_{\alpha}))/T]
\end{eqnarray}
For values of parameters and details see \cite{Tsang1}.

Here we do calculations for the production cross-sections of silicon
isotopes from projectile fragmentation of $^{48}$Ca and of copper
isotopes from projectile fragmentation of $^{86}$Kr using the canonical
thermodynamic model.  Some of these cross-sections are very small and
they serve as very stringent tests of the model. The model
has been extensively applied for production cross-sections of other particles
which are more numerous \cite{Das} and agreements are good.
The basic physics of the model is the same as in 
many other models of intermediate energy heavy ion collisions: the statistical 
multifragmentation model (SMM) \cite{Bondorf} or the microcanonical
simulations of heavy-ion collisions \cite{Gross,Randrup}.  But SMM or the microcanonical
simulations are totally impractical for calculations of very small
cross-sections  as they rely on Monte-Carlo
simulations.  The canonical model gives closed expressions and 
calculations can be made as accurate as desired.  The grand canonical
model is unsuitable for exploring these furthest limits of the phase-space
and is expected to be very unreliable \cite{Das,Das1}.  We will come back 
to this point later.

We will consider production of silicon isotopes from the statistical
break up of $^{48}_{20}$Ca.  We denote the avaerage number (multiplicity)
of $_{14}$Si$_n$ by $\langle n_{14,n}\rangle$.  Then the cross-section will
be $\sigma(14,n)=C\langle n_{14,n}\rangle$ where $C$ is a constant 
not calculable
from the thermodynamic model.  It depends upon dynamics which is 
outside the scope of the model.  Similarly we will consider the production
of Cu isotopes $_{29}$Cu$_{n}$ from a source which has 36 protons and 50
neutrons, i.e., $^{86}$Kr.  The source sizes adopted for the calculation are 
zero order guesses.  It could be sometimes smaller or greater 
depending on the diffusion from the target.

We will write down the formulae used for the calculations but we will
not derive them as they can be found elsewhere \cite{Das,Bhatta}.
The fragmenting system ($^{48}$Ca or $^{86}$Kr) has $Z_0$ protons
and $N_0$ neutrons.  The canonical partition function is given by
\begin{eqnarray}
Q_{Z_0,N_0}=\sum\prod \frac{\omega_{i,j}^{n_{i,j}}}{n_{i,j}!}
\end{eqnarray}
Here the sum is over all possible channels of break-up (the number of such
channels is enormous) which satisfy $Z_0=\sum i\times n_{i,j}$
and $N_0=\sum j\times n_{i,j}$; $\omega_{i,j}$ 
is the partition function of one composite with
proton number $i$ and neutron number $j$ respectively and $n_{i,j}$ is
the number of this composite in the given channel.
The one-body partition
function $\omega_{i,j}$ is a product of two parts: one arising from
the translational motion of the composite and another from the
intrinsic partition function of the composite:
\begin{eqnarray}
\omega_{i,j}=\frac{V_f}{h^3}(2\pi m(i+j)T)^{3/2}\times z_{i,j}(int)
\end{eqnarray}
Here $m(i+j)$ is the mass of the composite and
$V_f$ is the volume available for translational motion; $V_f$ will
be less than $V$, the volume to which the system has expanded at
break up. We use $V_f = V - V_0$ , where $V_0$ is the normal volume of  
nucleus with $Z_0$ protons and $N_0$ neutrons.  In this calculation we
have used a fairly typical value $V=6V_0$.

The probability of a given channel $P(\vec n_{i,j})\equiv P(n_{0,1},
n_{1,0},n_{1,1}......n_{i,j}.......)$ is given by
\begin{eqnarray}
P(\vec n_{i,j})=\frac{1}{Q_{Z_0,N_0}}\prod\frac{\omega_{i,j}^{n_{i,j}}}
{n_{i,j}!}
\end{eqnarray}
The average number of composites with $i$ protons and $j$ neutrons is
seen easily from the above equation to be
\begin{eqnarray}
\langle n_{i,j}\rangle=\omega_{i,j}\frac{Q_{Z_0-i,N_0-j}}{Q_{Z_0,N_0}}
\end{eqnarray}
The constraints $Z_0=\sum i\times n_{i,j}$ and $N_0=\sum j\times n_{i,j}$
can be used to obtain different looking but equivalent recursion relations
for partition functions.  For example
\begin{eqnarray}
Q_{Z_0,N_0}=\frac{1}{Z_0}\sum_{i,j}i\omega_{i,j}Q_{Z_0-i,N_0-j}
\end{eqnarray}
These recursion relations allow one to calculate $Q_{Z_0,N_0}$ 

We list now the properties of the composites used in this work.  The
proton and the neutron are fundamental building blocks 
thus $z_{1,0}(int)=z_{0,1}(int)=2$ 
where 2 takes care of the spin degeneracy.  For
deuteron, triton, $^3$He and $^4$He we use $z_{i,j}(int)=(2s_{i,j}+1)\exp(-
\beta e_{i,j}(gr))$ where $\beta=1/T, e_{i,j}(gr)$ is the ground state energy
of the composite and $(2s_{i,j}+1)$ is the experimental spin degeneracy
of the ground state.  Excited states for these very low mass
nuclei are not included.  For Si and Cu nuclei whose production 
cross-sections are sought in this work we use the experimental binding
energies tabulated in \cite{Audi} but also include a term for contribution
from excited states (see the discussion following).
For mass number $A=5$ and greater (but charge $\neq 14(29)$) we use
the liquid-drop formula.  For nuclei in isolation, this reads ($a=i+j$)
\begin{eqnarray}
z_{i,j}(int)=\exp\frac{1}{T}[W_0a-\sigma(T)a^{2/3}-\kappa\frac{i^2}{a^{1/3}}
-s\frac{(i-j)^2}{a}+\frac{T^2a}{\epsilon_0}]
\end{eqnarray}
The derivation of this equation is given in several places \cite{Bondorf,Das}
so we will not repeat the arguments here.  The expression includes the 
volume energy, the temperature dependent surface energy, the Coulomb
energy and the symmetry energy.  The term $\frac{T^2a}{\epsilon_0}$
represents contribution from excited states
since the composites are at a non-zero temperature.  
This form was used in other applications of
the model and we have kept this unchanged.  This term is
also included in $z_{14,n}(int)(z_{29,n}(int))$.  Note that in the 
fitting formula of Eq.(2) a different expression
for contribution from excited states is used. 

We have to state which nuclei are included in computing $Q_{Z_0,N_0}$.
For $i,j$, 
(the proton and the neutron number)
we include a ridge along the line of stability.  The liquid-drop
formula above also gives neutron and proton drip lines and 
the results shown here include all nuclei within the boundaries.

The long range Coulomb interaction between
different composites can be included in an approximation called
the Wigner-Seitz approximation.  We incorporate this following the
scheme set up in \cite{Bondorf}.  

It remains now to state the results.  Fig.1 taken from \cite{Mocko3}
shows the remarkable correlation between experimental values of cross-sections
of silicon isotopes (from $^{48}$Ca on $^9$Be reaction at 140 MeV per nucleon) and
average binding energies.  The experimental data on cross-sections
(shown as solid symbols in this paper)
span about seven orders of magnitude.  In Fig.2 we show results of our 
calculations(crosses).  There are basically two parameters: an overall
normalisation factor (chosen in the figure to give the correct value
of cross-section for $^{28}$Si) and the temperature (taken here to be
9.5 MeV which is within the range of temperatures expected for this
reaction).  Except at the tails of the distribution, 
the agreement is fair and the calculation
does indeed give the very rapid decrease of the cross-section for
large $A$.  The straggling in values of the cross-sections between
even-even and odd nuclei is also reflected in the calculation.  In Fig.3 
we compare data and calculations for the case of production of copper
isotopes.  The data here span more than eight decades  and the calculation,
except for the tails, does very well.  The straggling between 
cross-section values for odd-odd and odd copper isotopes is
highlighted in Fig.4.  In the same figure we show that both for data 
and calculation the straggling disappears if the cross-section is
plotted against $\langle B \rangle-\varepsilon_{pair}$ rather than
against just $\langle B \rangle$ (see also \cite{Tsang1}).  We find it 
gratifying that the model is able to reproduce such fine details.

Lastly, we will make a connection with grand canonical fitting of
the data \cite{Tsang1}.  In our model we use
$\sigma(z,n)=C\langle n_{z,n}\rangle$ where the value of
$C$ has to be taken from experiment.  
Thus in a model of this type what we need is the ratio
$\langle n_{z,n+1}\rangle/\langle n_{z,n}\rangle$ to be predicted
correctly.  From Eq.(6) this is
\begin{eqnarray}
\frac{\langle n_{z,n+1}\rangle}{\langle n_{z,n}\rangle}=
\frac{\omega_{z,n+1}}{\omega_{z,n}}\frac{Q_{Z_0-z,N_0-n-1}}
{Q_{Z_0-z,N_0-n}}
\end{eqnarray}
Here $Z_0,N_0$ are the charge and neutron number of the projectile which
is fragmenting ($^{86}$Kr or $^{48}$Ca) and $(z,n)$ denotes Cu or Si
isotopes.  The right hand side of eq.(9) is very simple in the grand
canonical ensemble.  The ratio of the two canonical partition functions
is replaced by a term independent of $n$, i.e., the right hand side is simply
$\frac{\omega_{z,n+1}}{\omega_{z,n}}\exp(\beta\mu_n)$
where $\mu_n$ is a constant for all $n,z$. 
That would be correct if $Z_0$ and $N_0$ are large and also
$Z_0\gg z$ and $N_0\gg n$ whereas in
small systems such as here, the ratio of the partition functions 
varies as $n$ changes.  As we approach neutron drip line it 
drops fast with $n$.  In order to approximate the right hand side
of eq.(9) by the simpler expression
$\frac{\omega_{z,n+1}}{\omega_{z,n}}\exp(\beta\mu_n)$ we need to choose
$\mu_n$ judiciously and further alter the value of the temperature
from the one used in the canonical model. In that case the temperature has 
to be significantly reduced.

In conclusion, the canonical model reproduces the salient features of
production cross-sections of very neutron rich nuclei.  The empirical
formulae for extrapolation \cite{Mocko3,Tsang1} are very useful
and the canonical thermodynamic model can not hope to replace these
but it aids to a theoretical understanding of the data.  The same 
parameters that we have used here can be used to predict the
production cross-sections of intrermediate mass fragments or the
properties of the largest fragment after mutifragmentation
\cite{Chaudhuri}.
\section{Acknowledgement} 
This work is supported by the Natural Sciences and Engineering Research 
Council of Canada and by the National Science Foundation under
Grant No PHY-0606007.

\begin{figure}
\includegraphics[width=5.5in,height=6.0in,clip]{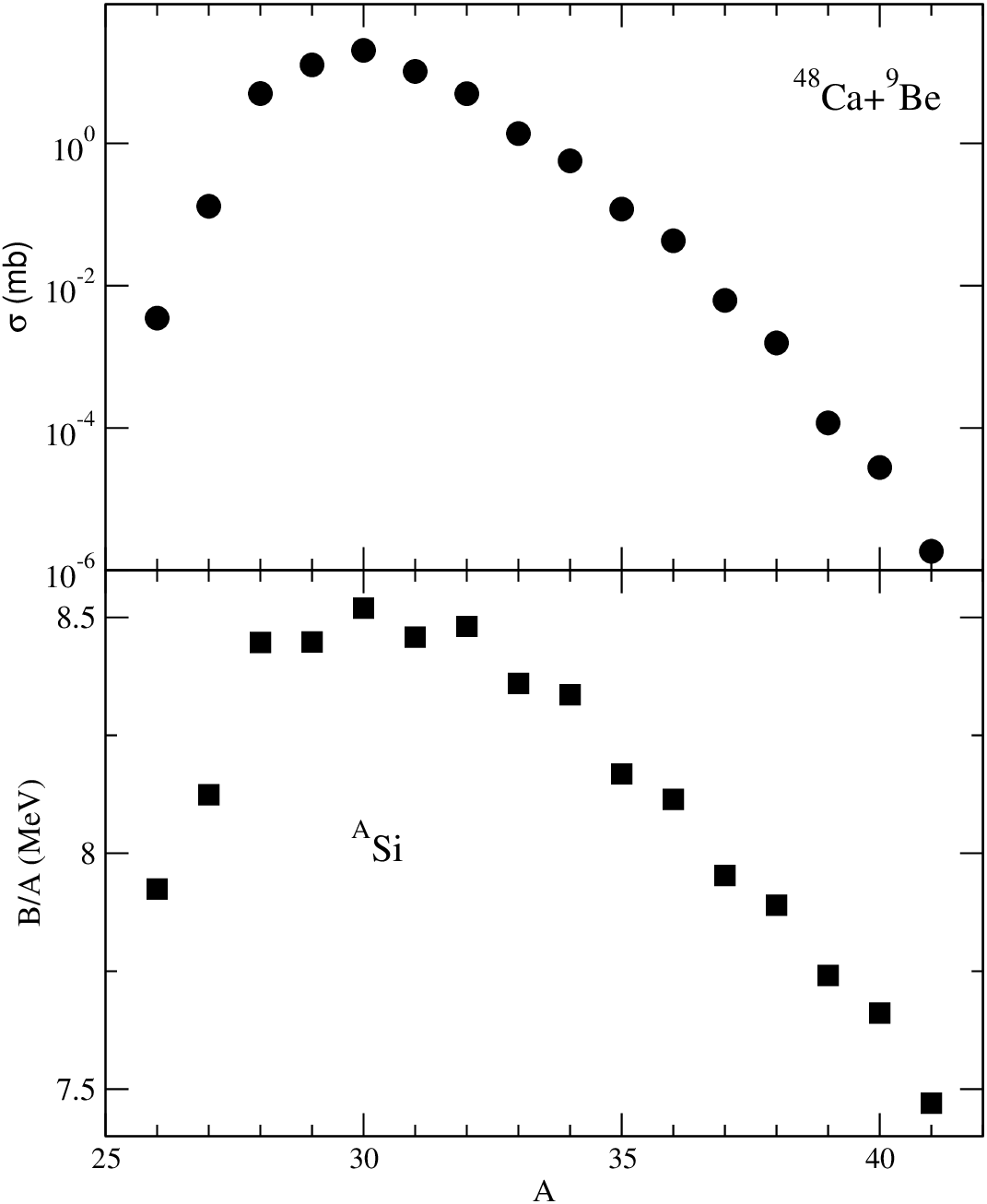}
\caption{ Fragment cross-section and average binding energy
plotted as a function of mass number for silicon isotopes.}
\end{figure}

\begin{figure}
\includegraphics[width=5.5in,height=5.0in,clip]{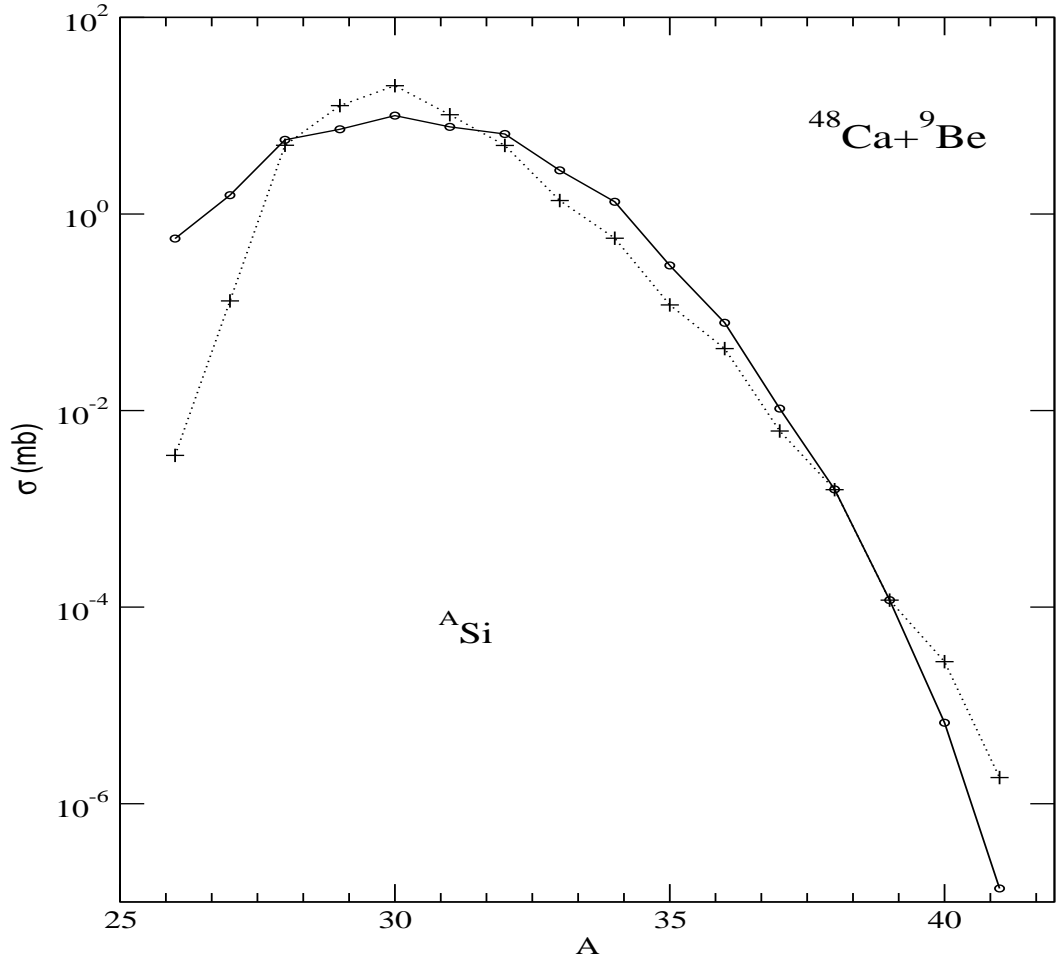}
\caption{ Experimental cross-sections for silicon isotopes (points)
compared with theoretical results (crosses).  A dotted line is
drawn through the experimental data and a solid line through the
calculated values.  The temperature used is 9.5 MeV.  Normalisation 
constant for theory is chosen by fitting to the experimental
cross-section of $^{39}$Si}.
\end{figure}

\begin{figure}
\includegraphics[width=5.5in,height=4.5in,clip]{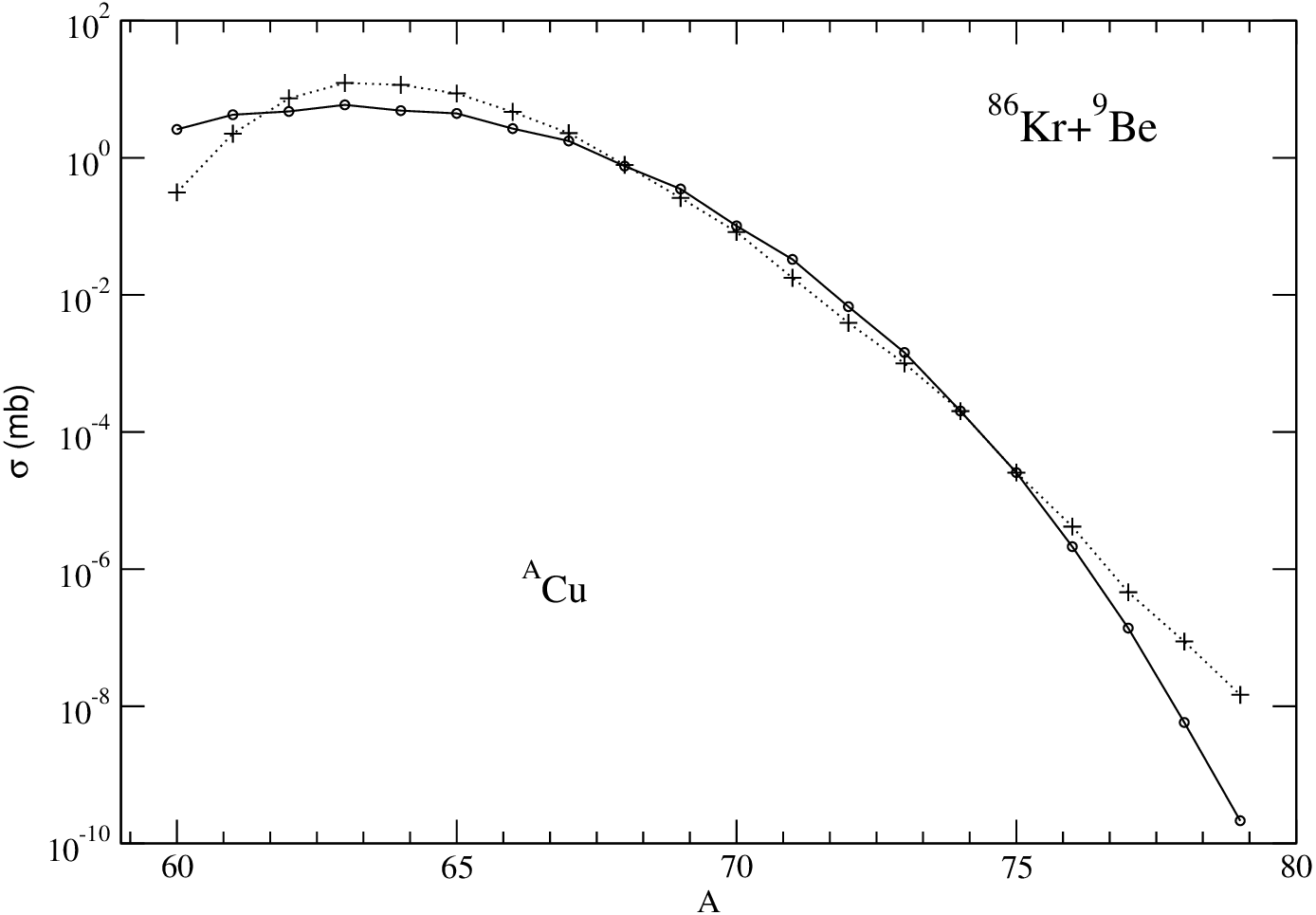}
\caption{ Same as in Fig.2 but for Cu isotopes. Normalisation
for theory is chosen from experimental value for $^{75}$Cu.}
\end{figure}

\begin{figure}
\includegraphics[width=5.5in,height=5.0in,clip]{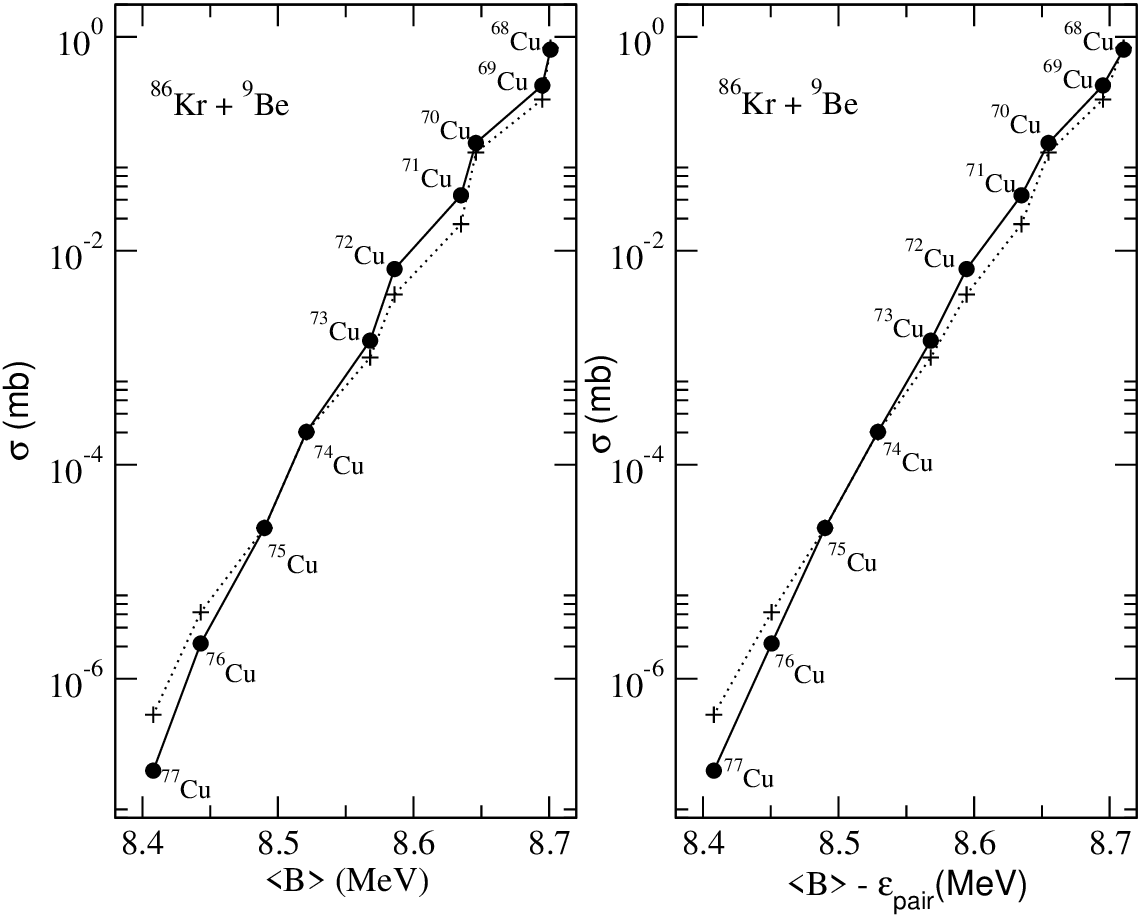}
\caption{ Straggling of data between odd-odd and odd-even cases
(points) when plotted against the average binding energy $<B>=B/A$.
Similar scatter is seen in theoretical calculation.  On the right
panel the cross-sections are plotted against $<B>-\varepsilon_{pair}$
(see eq.(1)).  The value of $\varepsilon$ used here is 30 MeV.  This 
decreases the straggling significantly both for data and theory.}
\end{figure}
\end{document}